\documentclass[12pt]{spieman}  
\usepackage{amsmath,amsfonts,amssymb}
\usepackage{graphicx}
\usepackage{setspace}
\usepackage{tocloft}
\usepackage[mathscr]{euscript}
\usepackage{lineno}

\title{Transition-edge sensor detectors for the Origins Space Telescope}

\author[a,*]{Peter C. Nagler}
\author[a]{John E. Sadleir}
\author[a]{Edward J. Wollack}
\affil[a]{NASA Goddard Space Flight Center, Greenbelt, Maryland, 20771 USA}

\cftpagenumbersoff{figure}
\cftpagenumbersoff{table} 
\begin{document} 
\maketitle

\begin{center}
{\small Accepted by the Journal of Astronomical Telescopes, Instruments, and Systems.}
\end{center}

\begin{abstract}
The Origins Space Telescope is one of four flagship missions under study for the 2020 Astrophysics Decadal Survey. With a 5.9 m cold (4.5 K) telescope deployed from space, {\sl Origins} promises unprecedented sensitivity in the near-, mid-, and far-infrared, from 2.8 -- 588 $\mu$m. This mandates the use of ultra-sensitive and stable detectors in all of the {\sl Origins} instruments. At the present, no known detectors can meet {\sl Origins'} stability requirements in the near- to mid-infrared, or its sensitivity requirements in the far-infrared. In this work, we discuss the applicability of transition-edge sensors, as both calorimeters and bolometers, to meet these requirements, and lay out a path toward improving the present state-of-the-art.
\end{abstract}


{\noindent \footnotesize\textbf{*}Corresponding author:  \linkable{peter.c.nagler@nasa.gov} }

\begin{spacing}{2}   

\section{Introduction}
\label{sec:intro}  

The Origins Space Telescope ({\sl Origins}) traces our cosmic history, from the formation of the first galaxies and the rise of metals to the development of habitable worlds and present-day life. {\sl Origins} does this through exquisite sensitivity to infrared radiation from ions, atoms, molecules, dust, water vapor and ice, and observations of extra-solar planetary atmospheres, protoplanetary disks, and large-area extragalactic fields. {\sl Origins} operates in the wavelength range 2.8 to 588 $\mu$m and is more than 1000 times more sensitive than its predecessors due to its large, cold (4.5 K) telescope and advanced instruments. 

A complete description of {\sl Origins}, its science goals, and its instrumentation is provided by the Origins Space Telescope Mission Concept Study Report\cite{Origins2019}. We reproduce some key details here. {\sl Origins} has three instruments, each with a variety of observing modes. The Origins Survey Spectrometer (OSS) is designed to perform extra-galactic surveys of far-infrared (FIR) line emission from 25 -- 588 $\mu$m, probing galaxy evolution out to $z\sim8.5$. OSS uses a series of grating modules to disperse incident light into six logarithmically-spaced bands, each with spectral resolving power $\mathscr{R} \sim 300$. A Fourier transform spectrometer (FTS) can be inserted into the optical path to achieve $\mathscr{R} \sim 40{,}000$, and the FTS can be used in conjunction with a scanning etalon to achieve $\mathscr{R} \sim 300{,}000$ in the 100 -- 200 $\mu$m band. In order to make background-limited observations of far-infrared spectral lines at $\mathscr{R} \sim 300$, OSS requires instrument sensitivity of $3.7\times10^{-21}$ $\mathrm{W/m^2}$ in an hour of observing at 200 $\mu$m \cite{Bradfordetal2020}. This corresponds to a required detector noise equivalent power (NEP) of $3\times10^{-20}$ $\mathrm{W/\sqrt{Hz}}$ and per-pixel saturation power of $\sim 0.02 - 0.2$ fW, a level of sensitivity that has yet to be achieved in a FIR detector. OSS has six focal planes, each with $\sim 10{,}000$ pixels.

The Far-Infrared Imager Polarimeter (FIP) instrument will perform wide-field photometric surveys of astrophysical objects, bridging the gab between observations by {\sl James Webb Space Telescope} ({\sl JWST}) and the Atacama Large Millimeter/sub-millimeter Array (ALMA). FIP has polarized imaging capability at both 50 $\mu$m and 250 $\mu$m. 
It requires $\sim 8{,}000$ pixels, a per-pixel NEP of $3\times10^{-19}$ $\mathrm{W/\sqrt{Hz}}$, and per-pixel saturation power of $\sim 2-20$ fW.

The Mid-Infrared Spectrometer Camera Transit spectrometer (MISC-T) will make spectral measurements of transiting exoplanets in the 2.8 -- 20 $\mu$m band with exquisite photometric precision, looking for biosignatures in the atmospheres of Earth-like exoplanets that orbit M stars. MISC-T uses a series of grisms to disperse light into three bands. The short wavelength bands (2.8 -- 5.5 $\mu$m and 5.5 -- 11 $\mu$m) will have $\mathscr{R} = 50 - 100$ and require $\sim 5$ ppm stability over 60 transits, and the long wavelength band (11 -- 20 $\mu$m) will have $\mathscr{R} = 165 - 295$ and requires $\sim 20$ ppm stability over 60 transits. Bright targets of K magnitude $\sim 3$ will be observed, leading to photon arrival rates of up to a few hundred thousand photons per pixel per second. 
Detector systems that meet these stringent stability requirements in the MISC-T band do not exist today.

The choice of detector used for each instrument will have tremendous impact on the development, construction, and eventual operation of {\sl Origins}. The sensitivity requirements of the FIR instruments mandate the use of sub-Kelvin detectors. There are several detector technologies that have the potential to meet the requirements of the FIR instruments. Among them are microwave kinetic inductance detectors (MKIDs), quantum capacitance detectors (QCDs), and transition-edge sensors (TESs), the subject of this paper. MKIDs and QCDs will be discussed elsewhere in this issue. The MISC-T instrument's stability requirement does not mandate the use of a low temperature detector. As a result the baseline detectors under consideration for MISC-T (HgCdTe arrays for wavelengths shorter than 10 $\mu$m  and Si:As arrays  for wavelengths longer than 10 $\mu$m) do not require sub-Kelvin coolers. They operate at $T \simeq 30$ K and $T \simeq 8$ K, respectively. Neither detector type has shown that it can meet the stability requirements of MISC-T, so in this paper we also present a TES calorimeter option for MISC-T. TES bolometers are also considered as candidates for MISC-T, but we argue that operation in a calorimetric mode provides an ideal choice.

This paper is organized as follows. In Section \ref{sec:TESs}, we introduce the TES and its most common implementations: calorimeters and bolometers that use a superconductor as a resistive thermometer. Then we present a path toward meeting the requirements of each {\sl Origins} instrument using TESs. In Section \ref{sec:calorimeters} we present the use of a photon-counting TES calorimeter for MISC-T. In Section \ref{sec:bolometers}, we discuss how TES bolometers can be designed to meet the requirements of OSS and FIP. Since no detector systems currently satisfy {\sl Origins}' requirements, we describe how TESs can. Our conclusions are in Section \ref{sec:conclusions}.  

\section{Transition-edge sensors}\label{sec:TESs}

\begin{figure}
    \centering
    \includegraphics[width=1.0\textwidth]{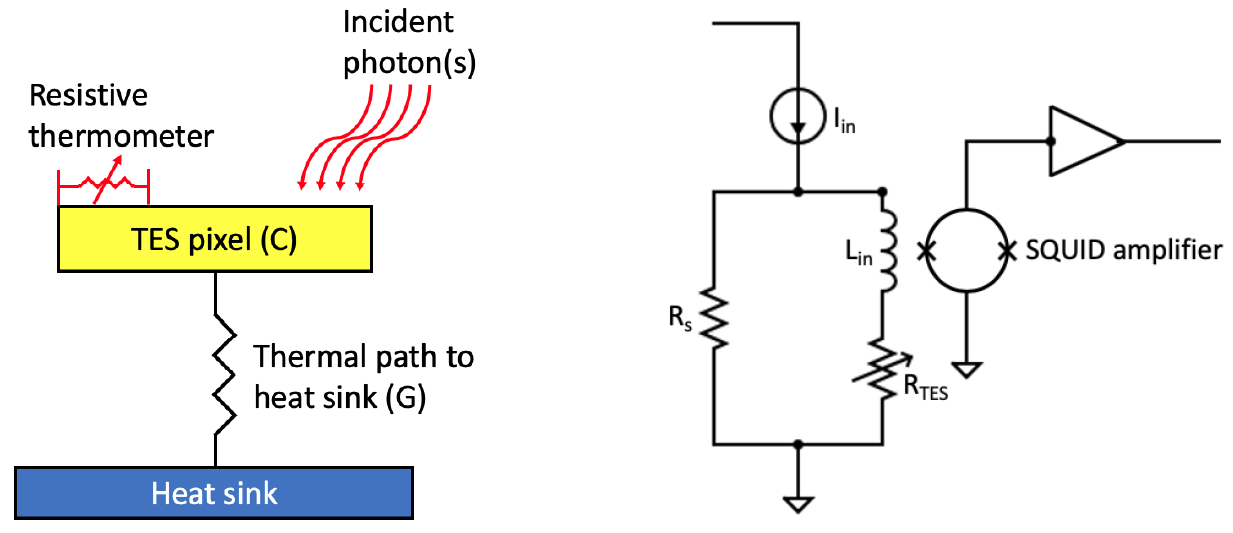}
    \caption{Left: Basic architecture of a TES detector. The variable resistor is a superconductor biased in its transition, employed as a resistive thermometer. The pixel has heat capacity $C$ and is isolated from the thermal bath via a thermal conductance $G$. Right: Basic TES readout circuit that uses a superconducting quantum interference device (SQUID) amplifier. The shunt resistor value $R_{\mathrm{s}}$ is chosen to be much smaller than the TES operating resistance $R_{\mathrm{TES}}$ in order to achieve voltage bias in the device. The SQUID amplifier is employed as an ammeter that measures changes in TES current due to absorption of energetic particles (calorimeter) or flux (bolometer).}
    \label{fig:TES}
\end{figure}

Transition-edge sensors detectors measure incident power or energy using the temperature dependence of a superconductor's resistance. First realized by Andrews in 1941\cite{Andrewsetal1941}, TESs are now the most sensitive detectors to radiation from the gamma-ray through the millimeter wave. 

TESs can operate as calorimeters or bolometers. A calorimeter is designed to measure discrete depositions of energy, and a bolometer is designed to measure quasi-static power dissipated by a flux of photons. Both implementations share the same basic architecture. Figure \ref{fig:TES} shows a cartoon of a TES and its basic readout circuit. The detector consists of a thermal mass of heat capacity $C$ at a temperature $T_{0}$, isolated from a thermal bath at temperature $T_{\mathrm{b}}$ across a thermal link with conductance $G$. The thermometer - a superconductor biased in its transition - measures the temperature of the isolated thermal mass, which can consist of electrons or electrons and phonons. The device is operated at near-constant voltage bias to stabilize the device in the superconducting transition via negative electrothermal feedback~\cite{Irwin1995}. A shunt resistor whose resistance is much smaller than the TES's at its operating point is typically used to achieve voltage bias. The shunt resistance value, in conjunction with the device and bias circuit inductance, sets the electrical time constant. 

The TES is readout/multiplexed out using one or more superconducting quantum interference device (SQUID) amplifiers. A range of SQUID multiplexing options currently exist for TES readout ({\it e.g.}, time-division multiplexing (TDM)~\cite{Irwin2002}, frequency-division multiplexing (FDM)~\cite{Irwin2002}, microwave SQUID multiplexing ($\mu$mux)~\cite{Irwin2004}, and code-division multiplexing (CDM)~\cite{Morgan2016}). Many TDM schemes are deployed in the field for reading out kilopixel-scale arrays of TES bolometers. Providing more bandwidth per pixel, microwave multiplexing shows the most promise for reading out large arrays of TES calorimeters; the Lynx mission is baselining arrays of $\sim 10^{5}$ TES calorimeters that are read out with $\mu$mux.  

For simplicity, the thermal conductance and heat capacity are commonly assumed to be independent of temperature over the range of operation, however, extension of this limiting case is necessary to model the full dynamic range and temporal response of the device in practical settings \cite{Irwin2005}. TESs are amendable to production entirely by available micro-fabrication techniques and are suitable for realization as large format detector arrays ({\it e.g.}, see for examples~\cite{Ullom_2015,Posada2015,Ade2015,Harper2018} {\it et cetera}). 

Both TES bolometers and TES calorimeters have achieved measured performance consistent with a linear near-equilibrium thermodynamics model of the system, with no hidden variables or unaccounted-for noise sources. This TES model is a powerful tool used to design detectors that meet the combined requirements for an application. It gives confidence in the ability to predict performance, help identify sources of unwanted characteristics, and make necessary changes that accelerate development programs.


\section{TES calorimeters}\label{sec:calorimeters}

TES calorimeters are designed to measure discrete depositions of energy. In this case, when an incident energetic particle of energy $E$ is absorbed by the TES, the temperature of the detector rises by $\Delta T = E/C$. This sudden temperature increase yields a corresponding increase of the superconductor's resistance, shunting current through the shunt resistor and reducing the current flowing through the SQUID input coil. The device then relaxes to its steady state temperature with an exponential decay time constant $\tau = C/G_{\mathrm{e}}$, where $G_{\mathrm{e}}$ accounts for electrothermal feedback gain \cite{Irwin1995}. This process yields a pulse in the time domain (Fig. \ref{fig:pulses}). In the X-ray band, where TES calorimeters are most commonly deployed, typical time constants are $\sim0.5$ ms. TES calorimeters designed for the near- to mid-infrared have time constants faster than $\sim 1$ $\mu$s\cite{Litaetal2008}. TES calorimeters are most commonly operated linearly where the pulse height is proportional to the deposited energy. In this way each TES pixel is a spectrometer, and an array of TES calorimeters is an imaging spectrometer on a chip ({\it i.e.,} an integral field spectrograph (IFS) that does not need dispersive optics). The energy scale of linear device operation is set by the heat capacity; the saturation energy -- the maximum deposited energy in the linear regime -- is defined as $E_{\mathrm{sat}} = C T_{0}/\alpha$, where $\alpha$ is the logarithmic temperature sensitivity of the device. With non-linear data analysis techniques, TES calorimeters can also be effectively operated non-linearly with little negative impact on device performance \cite{Fixsenetal2016,Buschetal2016}.

TES calorimeter development for astronomy applications has concentrated in development of X-ray calorimeter arrays. The X-ray work, along with complementary development of bolometers for microwave applications, has led to significant advancement in TES understanding and performance.  In this band TESs have achieved the highest resolving power of any non-dispersive spectrometer, measuring the energy of photons to better than a part in 3400.\cite{Miniussietal2018}. Some notable advancements in both TES understanding and performance include: 1) identification of sources of excess noise in TESs\cite{Luukanenetal2003,IrwinandHilton2005,Ullometal2004,Kinnunenetal2012,Iyomotoetal2008,Irwin2006}; 2) identification of the resistive mechanisms in TES sensors and ways to control the shape of the resistive transition surface\cite{Sadleiretal2010,Sadleiretal2011,Sadleiretal2013,Sadleiretal2014,Smithetal2013,Kozorezovetal2011,Kozorezovetal2011_2}; 3) improved understanding of energy losses from athermal phonons and quasiparticle excitations\cite{Kozorezovetal2013}; 4) improved fabrication methods and understanding of the thermal conductance of MEMs membranes and leg structures\cite{Holmesetal1998,Hoeversetal2005,Karvonenetal2010,Zenetal2014}; 5) improved coupling to radiation at longer wavelengths using tuned optical stacks\cite{Litaetal2008,Fukudaetal2011_2}; and 6) improved signal processing including methods for nonlinear signals\cite{Fixsenetal2002,Fixsenetal2004,Fixsenetal2016,Buschetal2016}.

Fundamental thermodynamic noise limits a TES calorimeter's achievable energy resolution $\Delta E_{\mathrm{FWHM}}$. For a TES, the known thermodynamic fluctuations are associated with electrical resistance (Johnson noise in the TES resistance $R_{0}$ and in the bias shunt resistor $R_{\mathrm{sh}}$) and thermal impedance (phonon noise across the thermal link $G$ that couples the sensor to the bath).   The expression for $\Delta E_{\mathrm{FWHM}}$ simplifies to a compact form \cite{IrwinandHilton2005} under the assumptions of negligible amplifier noise, negligible shunt resistor Johnson noise, and large loop gain:
\begin{equation}\label{eq:dE_full} 
\Delta E_{\mathrm{FWHM}} = 2 \sqrt{2 \log 2} \sqrt{\frac{4 k_{B} T_{0}^{2} C \sqrt{n/2}}{\sqrt{1-\left(T_{b}/T_{0}\right)^{n}}}\sqrt{\frac{1+2\beta}{\alpha^{2}}}}. 
\end{equation} 
Here $T_{0}$ and $T_{b}$ are the temperature of the TES and bath respectively, $n$ is a thermal exponent describing the power through the thermal link $G$, $C$ is the total heat capacity, and $\alpha$ and $\beta$ are both dimensionless parameters characterizing the sensitivity of the resistive transition to changes in temperature and current respectively.  More precisely, $\alpha$ and $\beta$ are defined as the logarithmic derivative of the resistance with respect to temperature and current, respectively:  $\alpha=\left(T/R\right)\times\left(\partial R/\partial T\right)$ and $\beta=\left(J/R\right)\times\left(\partial R/\partial J\right)$. The spectral resolving power $\mathscr{R}$ is the ratio of the photon energy to the energy resolution: $\mathscr{R} = E/\Delta E_{\mathrm{FWHM}}$ (for $\mathscr{R}$ of a few or more, $\mathscr{R} = E/\Delta E_{\mathrm{FWHM}} \simeq \lambda/\Delta \lambda_{\mathrm{FWHM}}$).  A detector with smaller $\Delta E_{\mathrm{FWHM}}$ is a higher resolution detector.    

Work by Bandler {\sl et al.}, 2008\cite{Bandleretal2008} and Iyomoto {\sl et al.}, 2008\cite{Iyomotoetal2008} are early examples that demonstrate a TES calorimeter's measured noise spectrum and energy resolution agree with those calculated from the model thermodynamic system (Eq. \ref{eq:dE_full}).  Assuming near-equilibrium thermodynamics and modeling the calorimeter system as one heat capacity $C$ connected through a single thermal conductance $G$ to a heat bath at temperature $T_{b}$, the calculated energy resolution departs from the measured value by less than 30\%.  If rather than the TES resistance being ohmic, it is assumed to be near-equilibrium with a linear current dependence, then the first order non-equilibrium correction term to the Johnson noise is shown to account for all the measured excess noise at the operating point.  
Measurements providing inputs to Eq. \ref{eq:dE_full} include impedance measurements and current-voltage (I-V) curves at different bath temperatures and bias currents, and measurements of the natural decay time $\tau$ of a pulse with no electrothermal feedback ($\tau=C/G$). Together with Eq. \ref{eq:dE_full}, these measurements predicted an energy resolution $\Delta E_{\mathrm{FWHM}} = 1.81$ eV. The measured energy resolution fitting to the Mn-k$\alpha$ complex at 5.9 keV agrees, giving $\Delta E_{\mathrm{FWHM}} = 1.81 \pm 0.10$ eV.

TES calorimeters have been designed to give high detection efficiency and constant energy resolution over a large spectral range.  In the visible, TES calorimeters exhibited uniform energy resolution from 0.3 eV to $> 10$ eV\cite{Romanietal1999}. TES calorimeters developed for quantum information science applications have achieved detection efficiencies of 99\%, the highest of any detector in the optical\cite{Litaetal2008,Fukudaetal2011_2}.

   
\subsection{Photon-counting TES for the mid-infrared}

In this section, we describe how a TES operated as a calorimeter, instead of a bolometer, can be used for the MISC-T instrument, reducing the requirements on instrument stability while maximizing effective throughput and overall observing time. 

\begin{figure}
    \centering
    \includegraphics[width = 1.0 \textwidth]{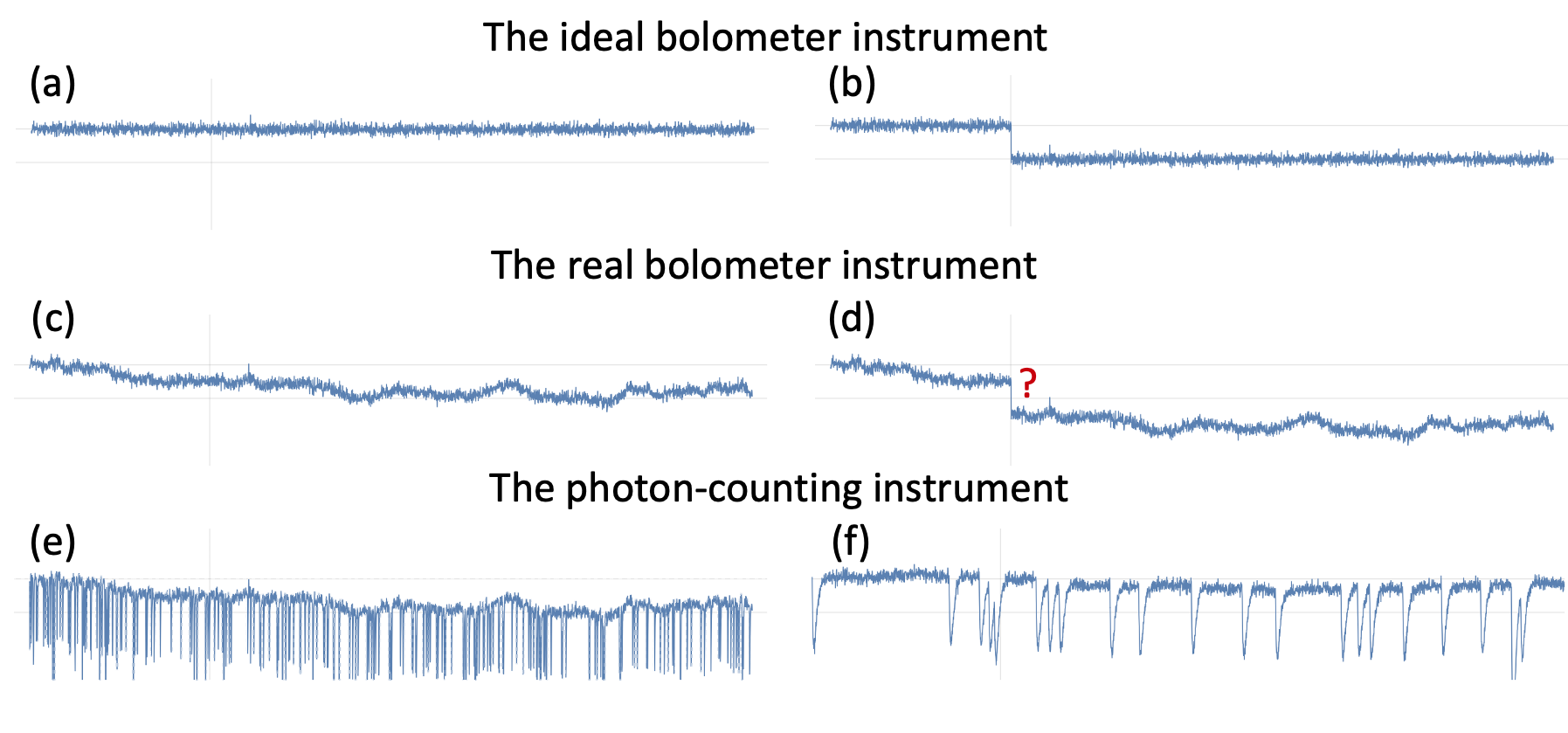}
    \caption{Illustration of the advantage to photon counting. The top frame shows how power is measured bolometrically with an ideal instrument. Incident optical power is manifested as a shift in the ``baseline'' TES current in the time domain. In the frequency domain, one would observe a higher white noise level due to incident photon noise. The center frame illustrates how system drift can be confused with an optical signal; the long time constants of bolometers place stringent requirements on system stability. The bottom frame shows how drift is mitigated if individual photon events are resolved (as opposed to just measuring a flux of photons.) In this case system drift has no effect on the measurement.}
    \label{fig:drift}
\end{figure}

A bolometer is unable to distinguish various external factors from the optical signal and therefore puts more stringent requirements on control of instrumental systematics, stability of the system, and the length of time over which such stability must be maintained. Compared to a calorimeter, a bolometer is more negatively impacted by system drift, stray power coupling into the detector, changes in ambient operating conditions including the local magnetic field environment, low frequency ($1/f$) noise, and array non-uniformities.  These effects manifest as a shift or drift in the baseline signal, the measured current flowing in the TES. Because the drift can be confused with an optical signal (Fig. \ref{fig:drift}), the burden to account for any drifts rests on the instrument as a whole. Especially as bolometers become more sensitive and therefore slower, commonly employed techniques to push the optical signal into a higher frequency band ({\sl e.g.}, by chopping) are more difficult to execute. By contrast, as long as the intrinsic resolution of the detector is high enough, this shift or drift in the baseline signal has negligible impact on a calorimeter's ability to detect a photon event. Moreover, photon-counting observations are more efficient. No external modulation is required and no observation time is spent observing a calibrator. From a sensitivity perspective, it has been shown in several previous studies that TES calorimeters can be realized with sufficient intrinsic energy sensitivity to count THz photons \cite{KarasikandSergeev2005,Weietal2008}.

\begin{figure}
    \centering
    \includegraphics[height = 5.2cm]{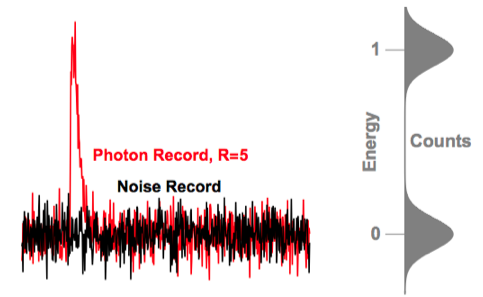}
    \caption{Example of our minimal criterion for noiseless single photon counting. Left: TES model simulated TES current time series records of a photon pulse with spectral resolving power $\mathscr{R}=5$ (red) and a noise record with no photon (black). Right: distribution of filtered energy of individual records with 1 or 0 photons.  Even with sensitivity of only $\mathscr{R}=5$ it is extremely unlikely to falsely report a 0-photon record (noise record) as an in-band photon.}
    \label{fig:counting}
\end{figure}

A TES dark count analog occurs when no photon is incident upon the detector and an anomalous noise trace triggers an event acquisition and application of the optimal filter to this noise record results in the energy of an in-band photon.  One finds that if you have a TES in its steady state operating bias collecting noise records at 50 MHz continuously in the dark with an energy resolving power of 4 for the lowest photon energy of the band of interest, the dark count event rate is of order one per age of the universe.  We use this $\mathscr{R}>4$ criterion for a ``noiseless'' detector (effectively no detector false positives). In Fig. \ref{fig:counting} we show an example for conditions that are just barely satisfying our criterion for noiseless photon counting with $\mathscr{R}=5$.  In red and black are TES current time series plotted for a 1 photon event with $\mathscr{R}=5$ (red) and a corresponding record for the TES at its operating point with a no photon record (black).  To the right is the distribution of many such records after application of an optimal filter to extract the energy of each record.  We see even for only $\mathscr{R}=5$ the population of 0 and 1 photon records are well separated.  The likelihood of mistaking a noise record for a photon record is extremely low.  This can also be seen as the 0 photon distribution at an energy of 1 is very small. As $\mathscr{R}$ increases beyond $\mathscr{R}=5$ the dark count rate remains negligible. Since the energy resolution $\Delta E_{\mathrm{FWHM}}$ is constant, as the photon energy increases the height of the pulse signal proportionally increases, leading to a proportionally-higher $\mathscr{R}$ measurement.
 
The energy sensitivity in the X-ray is used to determine the energy of an incident photon with high accuracy. In this application, the high energy sensitivity of a TES can be used to push noiseless single photon detection down to lower photon energies and into the MIR for {\sl Origins}.  The combination of high speed, noiseless single photon detection, and sensitivity are used to unambiguously distinguish in-band photon events from noise and identify and remove out-of-band events caused by cosmic rays or background. In grating-based dispersive spectrometers, the energy sensitivity also enables rejection of higher-order photons that correspond to a different spectral channel, a capability unique to energy-resolving single photon detectors and not found in other single photon technologies ({\sl e.g.,} electron multiplying charge coupled devices (EMCCDs)\cite{Zhangetal2009}, superconducting nanowire single photon detectors (SNSPDs)\cite{Vermaetal2019}, avalanche photo diodes (APDs)\cite{Petroffetal1987}, quantum dot detectors (QDs)\cite{Komiyama2011}, {\sl et cetera}.)

\begin{figure}
    \centering
    \includegraphics[height = 5.2cm]{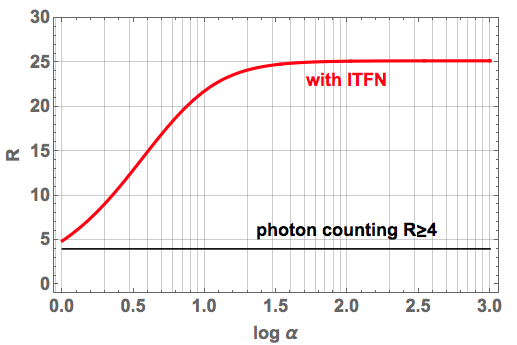}
    \caption{MISC-T TES spectral resolving power $\mathscr{R}$ at $\lambda = 20$ $\mu$m versus $\log \alpha$ operating at 70 mK. The TES model predicts this device designed for high absorption efficiency at $\lambda = 20$ $\mu$m gives sufficient energy sensitivity for noiseless photon counting with margin. Even very modest values of $\alpha$ give sufficient sensitivity to count photons with ample tolerance for system instabilities, which for realistic values only lead to minor degradation of energy resolution (see Chiao {\sl et al.}\cite{Chiaoetal2017})}
    \label{fig:Rvslogalpha}
\end{figure}

For the MISC-T instrument the TES calorimeter design for the lowest photon energy of the band ($\lambda = 20$ $\mu$m) is most challenging. It requires the lowest $\Delta E_{\mathrm{FWHM}}$ for noiseless photon counting, while also needing the largest absorbing area for an absorber-coupled strategy (see Section \ref{sec:radcouple}; $2 \lambda \times 2 \lambda$ is sufficient\cite{Chussetal2008,Thomasetal2010}). We therefore focus our discussion on the $\lambda = 20$ $\mu$m detector, which uses a $40 \times 40$ $\mu$m$^2$ resistive Bi absorber coupled to a TES sensor. We find applying the TES model to our MISC-T TES design achieves the required photon count rates, sensitivity, and absorption efficiency with known materials parameters. In Fig. \ref{fig:Rvslogalpha} we plot resolving power $\mathscr{R}$ for $\lambda = 20$ $\mu$m versus $\log \alpha$ over a range of typical $\alpha$ values.  We find that internal thermal fluctuation noise (ITFN) in the absorber is not negligible and departs from the simple analytic single body expression in Eq. \ref{eq:dE_full}.  The ITFN limits sensitivity and shows that increasing $\alpha$ much above 20 provides limited improvement in sensitivity.  The key take away is that even accounting for ITFN, the sensitivity is greater than the photon counting criterion (over 6 times larger).  This gives significant sensitivity margin even at the lowest photon energy of the MISC-T band.   Noiseless photon counting becomes easier (even greater margin) for the rest of the MISC-T band.

The thermal recovery time of TES calorimeters designed for the optical are typically of order 1 $\mu$s, accommodating count rates of $\sim$ 1 million counts per second.\cite{Fukuda_2009,Fukudaetal2011,Fukudaetal2011_2,Calkinsetal2011}  During the thermal recovery time, the TES is still able to receive and detect a photon, and therefore has no dead time, a feature unique among single photon detectors. If an additional absorption event occurs during the detector recovery signal pile-up occurs.  In such a case the ability to extract the energy of each event is degraded below the maximum resolving power of the detector.  For spectroscopy applications with no dispersive optics, such records are tagged as lower resolution events and different processing is employed to extract the event energies.  For MISC-T where the TESs are distributed over a spectral channel, pulse pile-up can be tolerable and the single photon events are identified by the fast rise of the leading edge provided the detector stays below the saturation energy.   Even for $\lambda = 20$ $\mu$m we see in Fig. \ref{fig:Rvslogalpha} the sensitivity is over 6 times larger than required to photon count and an increase in rate will not impact the ability to count photons.  The energy sensitivity margin increases further as the wavelength decreases.  As the rate increases further the calorimeter operation begins to become more bolometer-like and the advantages of photon counting start to diminish.

\begin{figure}
    \centering
    \includegraphics[width=1.0\textwidth]{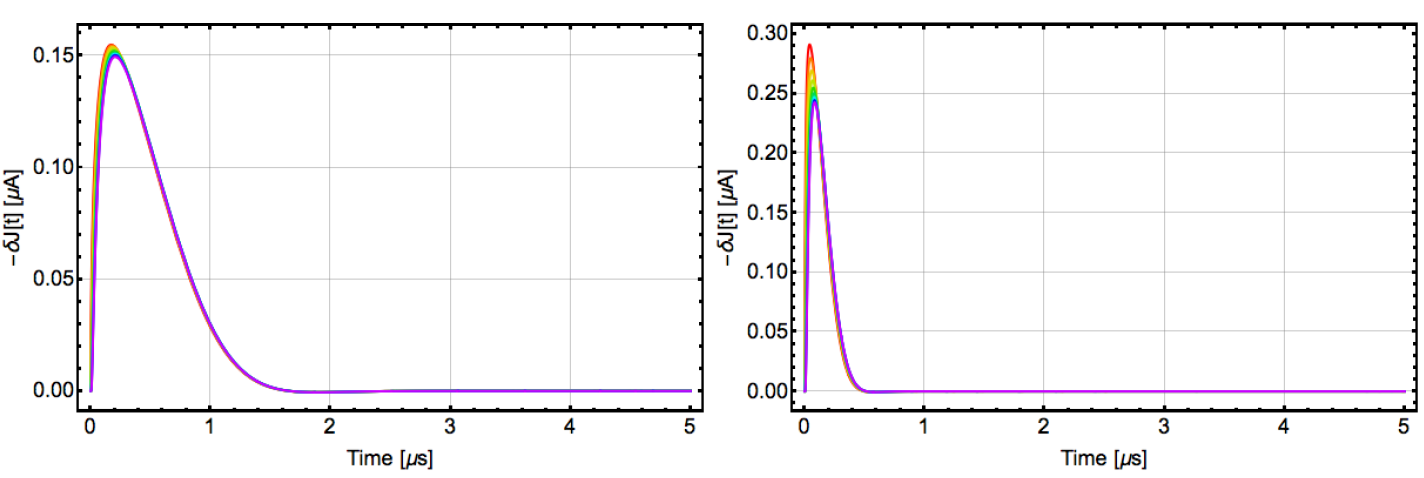}
    \caption{Simulated time domain signals of the MISC-T TES microcalorimeter response to $\lambda =$20 $\mu$m photons (60 meV). The bath temperature in this simulation is 50 mK, $\alpha = 10$, and the absorber consists of a 10 nm-thick Bi film. Left: Series of pulses absorbed at nine different points in a $40 \times 40$ $\mu$m$^{2}$ absorber at 70 mK. The 9 different pulses corresponds to 60 meV of energy deposited at 9 different positions in the absorber (increasing distance from TES with increasing hue).  The different colors show that there is some positional dependence to the pulse shape due to finite thermal conductance of the absorber structure. Right: A similar simulation, but at a 100 mK. This speeds up the detector response at the cost of energy resolution.}
    \label{fig:pulses}
\end{figure}

Increasing the number of spectral channels increases the spectral information of the measurement and decreases the photon rate per channel.  Increasing the event rate per spectral channel much above 1 million counts per second is achievable by : (1) decreasing the pulse recovery time; (2) increase the number of TES pixels per spectral channel (oversampling the point spread function); and (3) implementing photon counting algorithm for high count rates.  The faster pulse recovery time is achievable by operating the TES at higher temperatures (see Fig. \ref{fig:pulses} right) or by engineering a TES's thermal coupling.  Calkins {\sl et al.}\cite{Calkinsetal2011} have demonstrated engineering of the thermal coupling of the TES to achieve a thermal recovery time of less than half a microsecond.  In Fig. \ref{fig:pulses} we show the MISC-T TES response to a $\lambda = 20$ $\mu$m (60 meV) photon. In this TES model simulation 60 meV of energy is deposited at 9 different locations in a high  efficiency absorber.  The simulation on the left is at 70 mK and on the right 100 mK.  By raising the temperature we increase the speed of the detector. The faster pulses at higher temperatures show positional dependence of absorption which impacts the energy resolution. But even here the sensitivity is sufficient for noiseless photon counting with margin. The maximum count rate expected for MISC-T is $\sim 2.5 \times 10^{5}$ photons per second incident on the detector\cite{Origins2019}, thus we do not expect as-demonstrated device speeds to be limiting.

\section{TES bolometers}\label{sec:bolometers}

TES bolometers operate similarly to TES calorimeters, but as quasi-steady state devices, where power balance is achieved between sources of power dissipation (Joule heating of the sensor and incident radiation) and heat flow out the thermal link. Instead of being sensitive to single photon events, bolometers are sensitive to a quasi-steady flux of photons absorbed by the detector and manifested as deposited power. Bolometric operation of the TES has many of the traditional advantages of a transducer with large negative feedback including improved linearity, device speed, and immunity to parameter non-uniformity. When operated under negative electrothermal feedback \cite{Leeetal1995}, TES bolometers are highly linear devices. Thermal conductance to the bath sets the device saturation power and upper range of linear operation as a detector.

TES bolometers are the most sensitive detectors used to measure millimeter and sub-millimeter radiation. For applications requiring background-limited sensitivity, they are also the most widely deployed, with arrays of thousands of pixels operating from ground ({\it e.g.,} BICEP/Keck \cite{BICEP32016}), aircraft ({\it e.g.,} SOFIA/HAWC+ \cite{Harper2018}) and balloon ({\it e.g.,} SPIDER \cite{Bergmanetal2017} and EBEX\cite{EBEX2018}) platforms. Due to both their high sensitivity and the comparative ease of reading out large arrays, TES bolometers have largely replaced bolometers that use high-impedance thermistors as thermometers. At the present, however, no bolometers have demonstrated the ultimate sensitivity required for {\sl Origins}. In the following, we discuss what dictates a bolometer's sensitivity, the state-of-the art achieved, and some of the strategies being employed and natural tradeoffs that occur when designing a bolometer to meet {\sl Origins} requirements. These include thermal isolation techniques, radiation coupling techniques, and strategies to mitigate the impact of cosmic ray events.

\subsection{Bolometer sensitivity}\label{sec:sens}

A bolometer's sensitivity is parametrized by its NEP. In the dark, a bolometer's NEP is ideally limited by phonon noise ($N_{p}$), the fundamental thermodynamic noise that arises due to the random exchange of energy across the thermal link coupling the bolometer pixel to the thermal bath (see Fig. \ref{fig:TES}). In this limit, NEP $\simeq N_p$. Under optical bias, a background-limited detector is limited by photon noise, with $N_p$ as the leading sub-dominant noise term. Thus $N_p$ is the figure of merit for a bolometer's sensitivity (see Mather, 1982\cite{Mather1982} for a complete treatment of bolometer noise). The power spectral density of the phonon noise $\left|\left<N_p\right>\right|^{2}$, is given by

    \label{eq:Np}
\begin{equation}
    \left|\left<N_{p}\right>\right|^{2} = \gamma 4 k_{B} T_{0}^{2} G,
\end{equation}
where $k_{B}$ is Boltzmann's constant, $T_0$ is the temperature of the bolometer pixel, $G$ is the thermal conductance between the pixel and the thermal bath, and $\gamma$ is a constant that accounts for potential non-equilibrium effects. In the equilibrium case, where the pixel temperature $T_{0}$ is equal to the temperature of the thermal bath $T_b$, $\gamma = 1$. In the extreme non-equilibrium case, where $T_0 \gg T_b$, $\gamma = 1/2$. Bolometers are usually designed to operate somewhere between these extrema; e.g., bolometers designed for cosmic microwave background studies typically operate with $\gamma \simeq 0.62$. Thus to make a more sensitive detector, one must reduce the operating temperature and/or reduce the thermal conductance. The former must meet practical constraints of available cryostats (space qualified cryostats can achieve a minimum operating temperature of $\sim 50$ mK), so once a minimum operating temperature is reached, reductions in NEP are accomplished by reducing $G$.

NEP can be expressed in several ways that represent different measurements and device characteristics. It is important to note the distinctions between them. First is the thermal fluctuation noise NEP (NEP$_{\mathrm{TFN}}$). NEP$_{\mathrm{TFN}}$ is calculated from Eq. \ref{eq:Np} using a value for $G$ extracted from I-V curve measurements. It represents a theoretical limit where there is no significant contribution to the noise from other known or unknown noise sources (e.g., Johnson noise, amplifier noise, or excess noise). Next is the electrical NEP (NEP$_{\mathrm{el}}$). NEP$_{\mathrm{el}}$ is calculated from the measured current noise spectral density $i_{n}$ and frequency-dependent electrical responsivity $S_{\mathrm{el}}\left(\omega\right)$, where $S_{\mathrm{el}}\left(\omega\right) = S_{\mathrm{el}}/\left(1 + i \omega \tau\right)$\cite{Jones1953}. Here $S_{\mathrm{el}}$ is the DC electrical responsivity to dissipated Joule power, extracted from the I-V curve, and $\tau$ is the time constant of the device. Finally there is the optical NEP (NEP$_{\mathrm{opt}}$). In contrast to NEP$_{\mathrm{TFN}}$ and NEP$_{\mathrm{el}}$, NEP$_{\mathrm{opt}}$ is necessarily derived from optical measurements  and is calculated from the measured current noise spectral density $i_{n}$ and the device's responsivity to incident optical power $S_{\mathrm{opt}}$. In a background-limited detector NEP$_{\mathrm{opt}}$ would be dominated by photon noise. Both NEP$_{\mathrm{el}}$ and NEP$_{\mathrm{opt}}$ account for all noise sources, but only NEP$_{\mathrm{opt}}$ accounts for potential inefficiencies and loss mechanisms associated with photon absorption.

\begin{figure}
    \centering
    \includegraphics[width=1.0\textwidth]{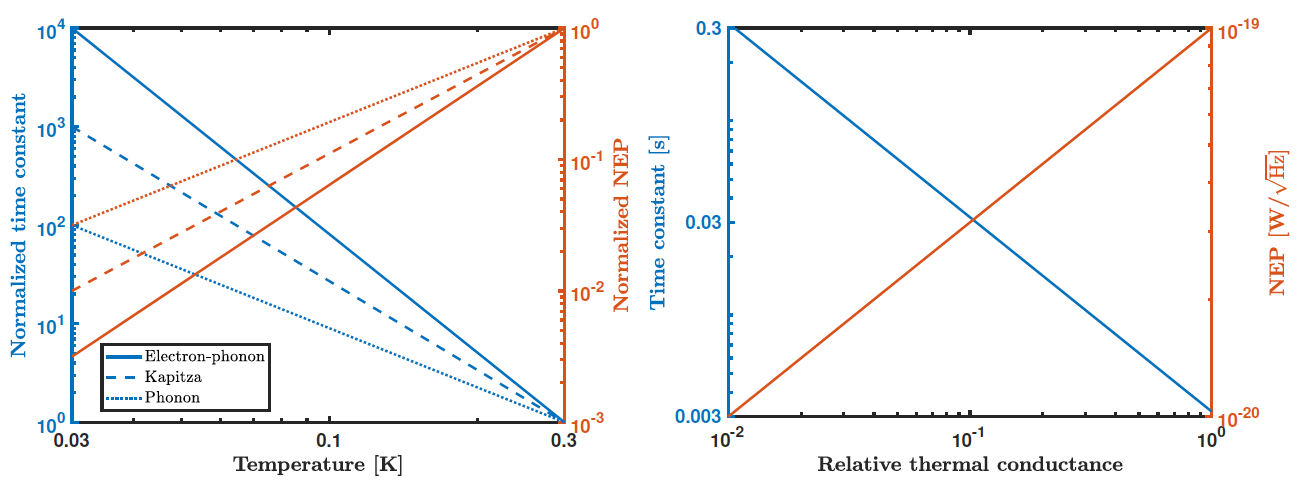}
    \caption{Left: Illustration of the relationship between a bolometer's operating temperature, time constant, and NEP when $G$ is dominated by electron-phonon conduction ($n=5$, solid), Kapitza conduction ($n=4$, dashed), and phonon conduction ($n=3$, dotted). The blue lines correspond to the axis on the left, and the red lines to the axis on the right. This plot assumes each bolometer has the same NEP and time constant at 300 mK. Real bolometers typically have exponents that take a value between 3 and 5, indicating multiple conduction mechanisms. Right: The impact on time constant if the most sensitive bolometer demonstrated to date\cite{Ridderetal2016} were modified to meet {\sl Origins} requirements by simply reducing $G$. To make the device the required 10x more sensitive requires a 100x reduction in $G$, which in turn makes the device 100x slower.}
    \label{fig:tau}
\end{figure}

Reductions in operating temperature $T$ and thermal conductance $G$ to achieve higher sensitivity both impact the bandwidth of a bolometer. Recall that the response time of a bolometer is proportional to $C/G$. At low temperatures, both $C$ and $G$ scale with temperature; $C \propto T$ and $G \propto T^{n}$. Bolometers are engineered to achieve a specific $n$ value depending on the physics of the heat transfer. Typical values include $n=3,4$, or $5$, corresponding to ballistic phonon conduction, Kapitza conduction, or electron-phonon conduction, respectively. In a realized bolometer, the measured $n$ value may take on intermediate values if there are multiple seres/parallel conduction paths through differing means or if there are few propagating modes\cite{Barrentineetal2018,Bartlettetal2019}. Figure \ref{fig:tau} illustrates the natural tradeoffs that occur between operating temperature, time constant, and NEP. In short, a more sensitive bolometer is also slower, thus introducing more stringent requirements on overall instrument stability. 

Engineering $G$ to achieve a certain value also impacts the saturation power of a bolometer $P_{\mathrm{sat}}$, defined as the power required to heat the bolometer to a temperature $T_{\mathrm{sat}}$ above which the responsivity of the device falls below some critical level. The relationship between $G$ and $P_{\mathrm{sat}}$ can be understood from the power balance requirement of steady-state operation, where the power flowing out the thermal link equals the sum of absorbed optical power and Joule power dissipated by the TES resistance. Note that stray power can manifest itself as either optical power or Joule power. For $G = AT^{n}$, where $A$ is a constant and $n$ is the exponent of the thermal conductance, $P_{\mathrm{sat}}$ is given by:
\begin{equation}
    P_{\mathrm{sat}} = \int_{T_{b}}^{T_{\mathrm{sat}}} G\left(T\right)dT = \frac{A}{n+1}\left(T_{\mathrm{sat}}^{n+1} - T_{b}^{n+1}\right).
\end{equation} 
Thus for a given bath temperature and sensor transition temperature ($T_{c}\simeq T_{\mathrm{sat}}$), the saturation power of the device scales with the thermal conductance. As $G$ is reduced to achieve higher sensitivity, the stray power requirements of the experimental platform -- electrical and optical -- become proportionally more stringent. In particular, implementation of techniques to mitigate stray power, like filtered connectors at the cryostat vacuum feedthroughs, thermal blocking and powder filters in close proximity to the detectors, and single-point grounding, are necessary to achieve a sufficiently quiet experimental space.

In the dark, the most sensitive bolometer demonstrated to date was developed by SRON and is described by Khosropanah {\sl et al.}\cite{Khosropanahetal2016}, Ridder {\sl et al.}\cite{Ridderetal2016}, and Suzuki {\sl et al.}\cite{Suzukietal2016}. It uses a TiAu TES on a leg-isolated SiN membrane, achieving NEP$_{\mathrm{TFN}}$ of $<1\times10^{-19}$ $\mathrm{W/\sqrt{Hz}}$ and NEP$_{\mathrm{el}}$ of $\sim 2\times10^{-19}$ $\mathrm{W/\sqrt{Hz}}$, with a phonon noise-limited bandwidth of $\sim 50$ Hz and a saturation power below a few fW. Figure \ref{fig:tau} illustrates the impact on this detector's time constant if $G$ were reduced by the necessary factor of $\sim 100$ to meet {\sl Origins} sensitivity requirements. Under optical bias, Karasik {\sl et al.} measured NEP$_{\mathrm{opt}}$ of $3 \times 10^{-19}$ $\mathrm{W/\sqrt{Hz}}$ using a Ti hot electron bolometer, the lowest optical NEP demonstrated to date in a TES bolometer \cite{KarasikandCantor2011}.

\subsection{Thermal isolation techniques}\label{sec:thermiso}

\begin{figure}
    \centering
    \includegraphics[width=1.0\textwidth]{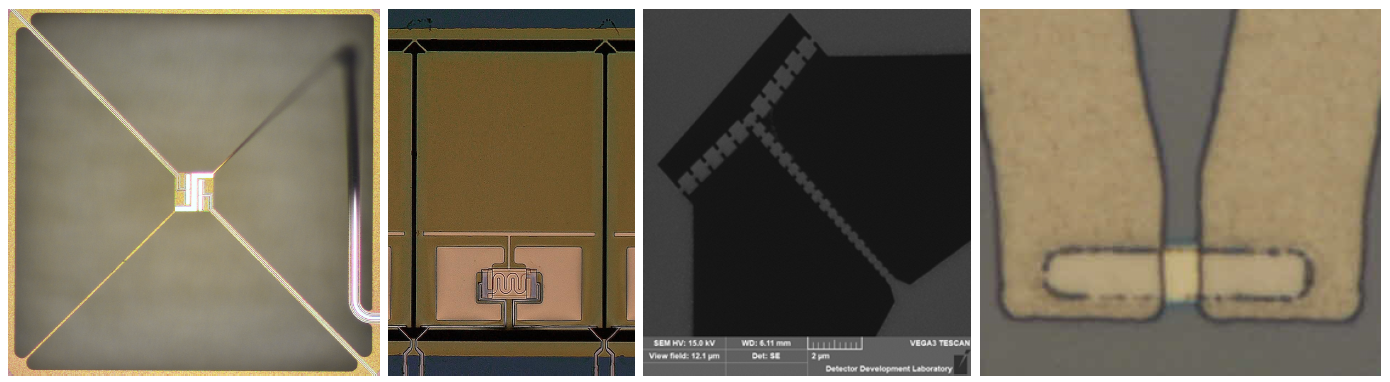}
    \caption{Examples of bolometers that use the thermal isolation techniques most likely to meet {\sl Origins} sensitivity requirements. Left: long-legged bolometer with micromachined SiN legs. Here the bolometer pixel is $\sim 50 \times 50$ $\mu$m$^{2}$, while the required per-pixel area is $\sim 500 \times 500$ $\mu$m$^{2}$. Devices like this have achieved excellent sensitivity, but suffer from a low fill fraction. Center left: Ballistic leg bolometer developed for the HIRMES instrument (image credit: A. Brown). Compared to the long-legged bolometer at left, the fill fraction offered by this type of bolometer is much higher. The membrane is micromachined from Si and is $\sim 680 \times 480$ $\mu$m$^{2}$ in area. Center right: Phononic leg supporting a bolometer membrane (image credit: K. Denis and K. Rostem). A realized phononic filter bolometer resembles the ballistic leg bolometer at center left, except nano-machined legs like the one pictured replace simpler dielectric beams. Right: Hot electron solid-substrate bolometer that accompishes thermal isolation via weak e-p coupling in the W sensor film. This device is $\sim 2 \times 2$ $\mu$m$^{2}$ in area. This device was designed for dark measurements and is not coupled to an antenna, however antenna coupling would be appropriate for optical characterization (see Section \ref{sec:radcouple}).}
    \label{fig:isolation}
\end{figure}

There are several thermal isolation techniques that show promise toward enabling {\sl Origins}-type sensitivities in bolometers. In no particular order, these fall into four basic categories (Fig. \ref{fig:isolation}): isolation via long (diffusive) legs, short (ballistic) legs, phononic filter/bandgap legs, and weak electron-phonon (e-p) coupling in the sensor material (a ``hot electron'' bolometer).

{\bf Long (diffusive) leg bolometers} generate thermal isolation via diffuse phonon transport across the  legs that support a membrane-isolated bolometer. The legs and the membrane are typically micromachined from Si or SiN. Conduction via diffuse phonon transport follows Fourier's law of thermal conduction: $G=\kappa A/\ell$, where $\kappa$ is the material's bulk thermal conductivity, $A$ is the cross sectional area of the leg, and $\ell$ is the length of the leg. As has been pointed out in the literature \cite{Rostemetal2014,Withingtonetal2017}, the Fourier limit is an oversimplification of the physics. $\kappa$ is not a bulk property, but rather a property that depends on the details of the leg's fabrication process, which can lead to non-uniform conductance across an array or between fabrication runs. In addition, the $G \propto \ell^{-1}$ behavior only holds for legs longer than a certain threshold; $\ell>400$ $\mu$m has been reported~\cite{Withingtonetal2017}. Bolometers that use long-leg isolation are exemplified in practice by the SPICA/SAFARI bolometers developed by SRON~\cite{Khosropanahetal2016,Suzukietal2016,Ridderetal2016} (see Section \ref{sec:sens}). 
A representative optical micrograph of a long-leg bolometer is shown in Fig. \ref{fig:isolation}. Despite the excellent achieved sensitivity of long-legged bolometers, an obvious limitation is the low fill fraction achieved. By area, the SPICA/SARARI bolometers have a fill fraction of $\sim 4\%$. SPICA/SAFARI has three bands with between 600 and 2000 pixels per band, yielding far fewer pixels than the $\sim 60{,}000$ required for OSS. With $G\propto A/\ell$, reducing the conductance by 100x to meet OSS requirements would necessarily reduce the fill fraction. For FIP, however, these bolometers already meet the sensitivity requirement and the arrays size is similar to SPICA/SAFARI. An added complication to longer legs is that as the leg length increases, the heat capacity of the legs can become non-negligible, leading to measurable ITFN degrading device performance. Long-legged bolometers lend themselves most naturally to resistive absorber coupling. Antenna coupling would likely have to incorporate the legs themselves, and further investigation is required to determine whether the electromagnetic requirements of an antenna could be consistent with the thermal requirements of the legs.

{\bf Short (ballistic) leg bolometers} achieve thermal isolation by minimizing the solid angle between the radiating sensor and legs that support a membrane. The ballistic limit occurs when the length of the leg is shorter than the phonon mean free path $\ell_{\mathrm{mfp}}$ in the material, so it is assumed that phonons that pass through the leg aperture escape to the bath. For very small cross sectional areas, ballistic legs have been shown to approach the so-called ``quantum limit'' of thermal conductance, where only four phonon modes can propagate\cite{Osmanetal2014}. The bolometers developed for the HIRMES instrument\cite{Barrentineetal2018} used this technique to achieve kilopixel arrays with NEP$_\mathrm{el}$ $<1.8 \times 10^{-18}$ $\mathrm{W/\sqrt{Hz}}$\cite{Brown2020}, in good agreement with the targeted value of NEP $<3 \times 10^{-18}$ $\mathrm{W/\sqrt{Hz}}$ for background-limited operation. Compared to long-legged bolometers, ballistic leg bolometers can achieve much higher array fill fractions, and the achieved thermal conductance of realized legs is both calculable from real bulk material properties and is much more tolerant of fabrication non-idealities and non-uniformities~\cite{Osmanetal2014,Rostemetal2016,Williamsetal2018}.

{\bf Phononic filter bolometers} use nano-machined legs to generate coherent phonon scattering that suppresses propagated phonon modes to below the quantum limit.\cite{Clelandetal2001,Zenetal2014,Rostemetal2016,Williamsetal2018}. Like ballistic leg bolometers, phononic filter bolometers promise short legs that enable high array filling fractions, but with the additional advantage of reduced thermal conductivity relative to ballistic leg designs. Theoretical work by Rostem {\sl et al.}\cite{Rostemetal2016} indicates that a factor of 5 improvement over the quantum limit is possible with phononic filter legs that are just 10 $\mu$m long and are machinable using known techniques. Devices presented by Williams {\sl et al.}\cite{Williamsetal2018} achieved a factor of $\sim 2$ reduction in thermal conductance below the quantum limit using phononic filter legs. Fabrication of phononic filters is challenging as sub-$\mu$m lithography techniques are required, but recent work by Denis {\sl et al.}\cite{Denisetal2019} demonstrates reliable and robust fabrication of phononic leg-isolated TESs.

{\bf Hot electron bolometers (HEBs)} use the inherent decoupling of electrons and phonons in certain metals at low temperatures for thermal isolation. The conductance between electrons and phonons at low temperature is given by
\begin{equation}
G_{ep} = \Sigma V T^{n},
\end{equation}
where $\Sigma$ is the electron-phonon coupling constant (material property), $V$ is the volume of the sensor, $T$ is temperature, and $n$ is the thermal exponent (typically $n=5$ for e-p coupling). A table of measured coupling constants for superconducting metals that would act as the hot electron bolometer is provided by Karasik {\sl et al.}\cite{Karasiketal2011} Among superconductors, W has the weakest known e-p coupling and therefore the greatest potential as a hot electron bolometer sensor material. Karasik {\sl et al.}\cite{Karasicetal2010,Karasiketal2011,KarasikandCantor2011} have achieved NEP$_{\mathrm{el}}$ below $1 \times 10^{-19}$ $\mathrm{W/\sqrt{Hz}}$ and NEP$_{\mathrm{opt}}$ of $3 \times 10^{-19}$ $\mathrm{W/\sqrt{Hz}}$ using Ti as the hot electron sensor; we have performed simulations that indicate W-based HEBs can exceed OSS requirements at 70 mK. A significant advantage of HEBs over the other bolometer types described above is that HEBs do not require any membranes, thus fabrication is simplified, large fill fractions can be realized, and the final devices are mechanically robust. The fabrication problems that can occur in HEBs ({\sl e.g.}, inconsistent transition temperatures or interface problems) are shared among all TES bolometer types and ultimately are not limiting. Note that while there are comparatively few examples of realized hot electron bolometers in the literature, calorimeters are frequently designed to operate in the electron-phonon decoupled limit using materials that would also yield exceptional sensitivity in a bolometer ({\sl e.g.}, see Calkins {\sl et al.}, 2011\cite{Calkinsetal2011}).

\subsection{Radiation coupling techniques}\label{sec:radcouple}

Antenna-coupling can be used to define the sensor's angular acceptance and coupling to the electromagnetic radiation field. It converts the incident fields (photons) into electronic excitations in the absorber media which are dissipated and subsequently detected as heat. To achieve high antenna coupling efficiency requires transforming the modal symmetry and impedance scales encountered by the wave in free space to that present in the circuit elements employed to absorb the radiation. A wide variety of techniques exist to carry out these functions; however, due to the need to realize large arrays of sensors -- in particular -- methods amendable to implementation as planar lithographic structures are primarily of interest. Consideration of the overlap in the angular response of the individual antenna elements in a multi-beam array can provide insights into the inter-pixel isolation and maximum achievable power coupling~\cite{Stein1962}. Similarly, care is warranted in providing isolation between the detector bias/readout and the optical signal paths.  Rejection of out-of-band radiation sources, which can potentially load a bolometric sensor and degrade the achievable response, also needs to be realized. These functions are typically achieved through the use of choke circuits on the sensor wafer and cooled thermal blocking filters at appropriate locations in the instrument system. 
 
\begin{figure}[!t]
	\centering
	\includegraphics[width=0.9\textwidth]{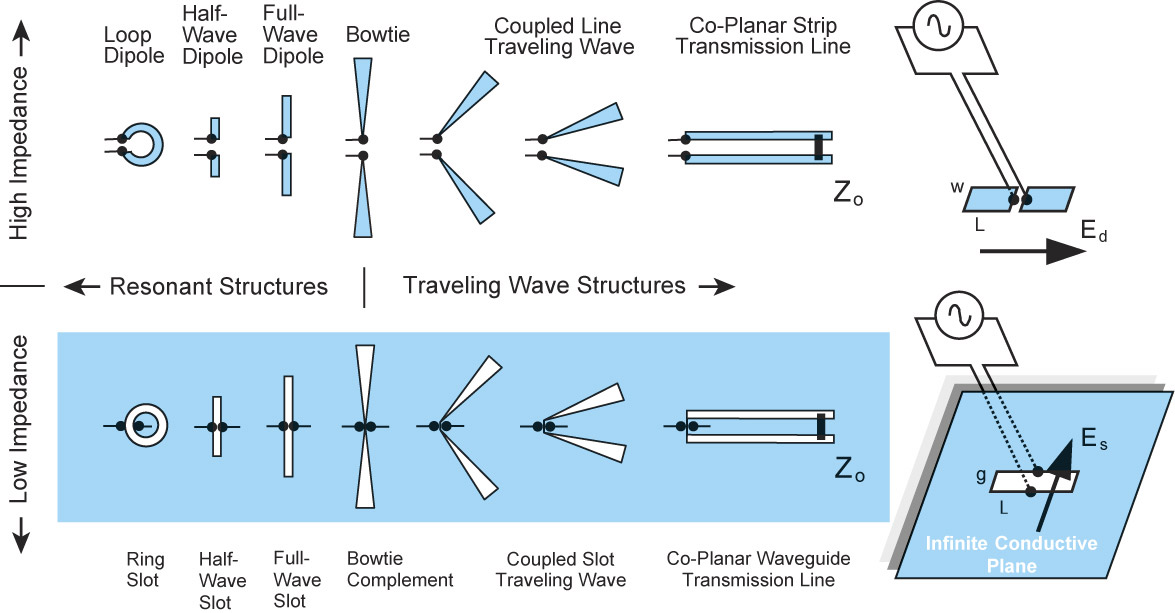}
	\caption{Planar slot aerials and their relation to complementary wire aerials (Babinet's principle)~\cite{Booker1946}.}
	\label{figure:antennas}
	\vspace{-0.1in}
\end{figure}

A qualitative survey of commonly encountered planar antennas such as resonant and traveling-wave line- and slot-structures is provided in Fig.~\ref{figure:antennas}. The angular response (or ``antenna pattern'') as well as the impedance scales of complementary line- and slot-like aerial structures are linked by Babinet's principle in its vector form~\cite{Booker1946}. Extension of these symmetry concepts to antenna geometries which are only a function of angle enables the realization of frequency independent antenna structures~\cite{Rumsey1966}. Examples of frequency-independent traveling-wave structures include planar logarithmic spiral and log-periodic antennas, which have continuous and discrete scaling symmetries respectively. In practice, such antennas exhibit deviations from frequency independence from truncation at the inner and outer length scales ({\it i.e.,} set by the feed network and and the maximum extent of the aerials). While this tends to simplify impedance matching to the detector's absorber circuit, these structures tend to have a low filling-fraction when used in arrays over a wide spectral range. The operation of planar antenna arrays in an immersion lens configuration~\cite{Rutledge1982,Rutledge1983,Compton1987} has enabled greater control over electromagnetic substrate losses and has allowed a natural separation of the radiation and readout functions of the focal plane. 

While antenna coupled arrays have found utility in instrument applications where high spatial-sampling presents a driving consideration~\cite{Griffin2002}, an alternative approach to achieve full sampling is presented by ``absorber coupled'' sensor arrays~\cite{Kusaka2014}. In this limit, the ``antenna response'' of the sensor is essentially uniform over the range of interest and the telescope optics in concert with a cold Lyot stop is used to specify and limit radiation presented to the focal plane~\cite{Wollack2006}. In this configuration a homogeneous thin film for each pixel's absorber has resonant (quarterwave) or frequency independent back-termination~\cite{Clarke1977,Carli1981} depending on the desired spectral range and coupling. Incorporation of frequency selective surfaces (FSS) in the absorber structure~\cite{Kowitt1996,Monacelli2005} provides an alternative path to achieving the desired impedance scales for high in-band-coupling and improved out of band rejection in bolometric sensor settings.

Prior discussions of antenna coupling for bolometer implementations can be found in the literature~\cite{Schwarz1977,Mees1991,Karasiketal2011}. It is important to note that for a bolometric sensor, the detailed thermal mode of operation can place practical constraints on how the device is electromagnetically coupled to the radiation field. For example, in hot-electron bolometers, the antenna has direct electronic coupling and appropriate material selection enables operation of the device. Similarly, in phononic filter, ballistic leg, and diffusive leg bolometers, direct electrical contact would thermally short the detector and reactively electromagnetically coupling the antenna to a resistive absorber element on a thermally isolated element is typically employed. Either approach can be employed for calorimeter applications, though the device we present in Section \ref{sec:calorimeters} uses a resistive absorber element. In all cases, the choice of absorber strategy should reflect that array fill fraction is a parameter which should be maximized in large array applications like {\sl Origins}. For completeness, related antenna-coupled sensors at infrared wavelengths~\cite{Wang1975} employing other physical detection mechanisms are also conveyed. These non-cryogenic implementations have included integrated dipole antennas~\cite{Fumeaux2000}, bowtie antennas~\cite{Chong1997,Gonzalez2005}, log-periodic/spiral antennas~\cite{Chong1997}, microstrip patch antennas~\cite{Codreanu1999}, microstrip dipole antennas~\cite{Codreanu2001}, and Yagi-Uda arrays~\cite{Mohammadi2020}.

\subsection{Susceptibility to cosmic rays}\label{sec:cosmicrays}

The interaction between energetic charged particles and the materials used in the detector system lead to a stochastic background of energetic events observed by the sensors in the focal plane over the course of a space mission. Consider for example the {\sl Planck Surveyor} mission which reported a rate of 80 cosmic ray related events per minute for the High Frequency Instrument array during operations in its L2 orbit. Even with template fitting of its radiometric data, a science data loss of $\sim 10-15\%$ was experienced due to these ``glitch''-like temporal events~\cite{PlanckIV, PlanckVI}. Beyond impacting observational efficiency, non-ideal detector responses associated with these energetic events can lead to instrumental stability and calibration issues which reduce imaging fidelity if unmitigated. {\sl Origins} will require higher sensitivity and by extension a lower focal plane operating temperature than used in previously deployed systems and thus increase the relative importance of the sensor's thermal bus implementation on minimizing the impact of cosmic rays. Extension of the detector design techniques employed for calorimetry~\cite{Saab2004,Kilbourne2006,Iyomoto2009,Finkbeiner2011} provide a viable path to address this instrumentation need. Mitigating cosmic rays is one particular advantage of a device known as the ideal integrating bolometer (IIB)\cite{Kogut2002,Kogutetal2004,Nagleretal2016,Canavanetal2017}, which combines leg isolation (diffusive, ballistic, or phononic) with a switchable thermal short. Upon an upset ({\it e.g.,} cosmic ray hit), the device can be reset quickly to mitigate the data losses that {\sl Planck} experienced.

\section{Conclusions}\label{sec:conclusions}

The unprecedented sensitivity enabled by {\sl Origins}' large, cold, and space-based telescope places stringent requirements on its instruments. In particular, the detector systems employed by each instrument must be designed to enable background-limited observations, contributing negligible noise to the overall instrumental budget. Detectors that enable background-limited observations in each channel do not exist today. For MISC-T, the most difficult specification the detector system must meet is stability. While bolometers and semiconducting detectors have many sources of instability ({\sl e.g.}, environmental drifts for bolometers, dark current and read noise for semiconductors), all these terms go to zero if a TES calorimeter is employed. The TES calorimeter we designed using well-established physical models and measured material parameters can overcome many of the other challenges associated with the MISC-T detector system, such as high photon flux and the large required bandwidth, delivering all the advantages of photon-counting using an energy-resolving detector. For OSS and FIP, the most challenging requirement the detector subsystems must meet is sensitivity, with radiation coupling and cosmic ray immunity also leading design drivers. There are multiple promising paths toward achieving the required sensitivity using TES bolometers, with demonstrated sensitivities now approaching the OSS requirement and improving. For all the instruments, the TES detector options benefit from strong heritage in astrophysics instrumentation and well-understood physics that enables the design and implementation of devices that operate at the thermodynamic limit across the {\sl Origins} band.

\subsection*{Disclosures}
The authors declare no conflicts of interest.

\subsection* {Acknowledgments}
Research supported by the National Aeronautics and Space Administration (NASA) under the Goddard Space Flight Center (GSFC) Internal Research and Development (IRAD) program. The authors are grateful for figures and technical feedback provided by Ari Brown, Kevin Denis, and Karwan Rostem of the GSFC. We are also grateful for the feedback provided by the referees, which improved the quality of this paper.


\bibliography{origins_tes}   
\bibliographystyle{spiejour}   


\vspace{2ex}\noindent\textbf{Peter C. Nagler} is an experimental physicist in the Instrument Systems and Technology Division at NASA's Goddard Space Flight Center. His research focuses on low temperature detectors and custom instrumentation for astrophysics applications.
\vspace{12pt}\\
\noindent\textbf{John E. Sadleir} is a condensed matter physicist in the Detector Systems Branch at NASA's Goddard Space Flight Center. His research focuses on cryogenic detectors for particle physics, cosmology, and astrophysics applications.
\vspace{12pt}\\
\noindent\textbf{Edward J. Wollack} is a research astrophysicist in the Observational Cosmology Laboratory at NASA's Goddard Space Flight Center. His interests include cosmology, astronomical instrumentation, and electromagnetics.

\listoffigures
\listoftables

\end{spacing}
\end{document}